\documentclass[12pt]{article}
\usepackage{graphicx,amssymb,amsfonts,amsmath,epsfig,latexsym}

\makeatletter \@addtoreset{equation}{section} \makeatother

\begin{document}
\makeatletter
\def\flE{\begin{picture}(0,0)
   \put( 0.25,    0){\vector( 1, 0){0.50}}
   \@ifstar{\@flE}{\@@flE}}
\def\@flE  #1{\put( 0.5 ,-0.03){\makebox(0,0)[ t]{$#1$}}\end{picture}}
\def\@@flE #1{\put( 0.5 , 0.03){\makebox(0,0)[ b]{$#1$}}\end{picture}}
\def\flNE{\begin{picture}(0,0)
   \put( 0.18, 0.18){\vector( 1, 1){0.64}}
   \@ifstar{\@flNE}{\@@flNE}}
\def\@flNE #1{\put( 0.52, 0.48){\makebox(0,0)[tl]{$#1$}}\end{picture}}
\def\@@flNE#1{\put( 0.48, 0.52){\makebox(0,0)[br]{$#1$}}\end{picture}}
\def\flN{\begin{picture}(0,0)
   \put(    0, 0.20){\vector( 0, 1){0.60}}
   \@ifstar{\@flN}{\@@flN}}
\def\@flN  #1{\put( 0.03, 0.5 ){\makebox(0,0)[ l]{$#1$}}\end{picture}}
\def\@@flN #1{\put(-0.03, 0.5 ){\makebox(0,0)[ r]{$#1$}}\end{picture}}
\def\flNW{\begin{picture}(0,0)
   \put(-0.18, 0.18){\vector(-1, 1){0.64}}
   \@ifstar{\@flNW}{\@@flNW}}
\def\@flNW #1{\put(-0.48, 0.52){\makebox(0,0)[bl]{$#1$}}\end{picture}}
\def\@@flNW#1{\put(-0.52, 0.48){\makebox(0,0)[tr]{$#1$}}\end{picture}}
\def\flW{\begin{picture}(0,0)
   \put(-0.25,    0){\vector(-1, 0){0.50}}
   \@ifstar{\@flW}{\@@flW}}
\def\@flW  #1{\put(-0.5 , 0.03){\makebox(0,0)[ b]{$#1$}}\end{picture}}
\def\@@flW #1{\put(-0.5 ,-0.03){\makebox(0,0)[ t]{$#1$}}\end{picture}}
\def\flSW{\begin{picture}(0,0)
   \put(-0.18,-0.18){\vector(-1,-1){0.64}}
   \@ifstar{\@flSW}{\@@flSW}}
\def\@flSW #1{\put(-0.52,-0.48){\makebox(0,0)[br]{$#1$}}\end{picture}}
\def\@@flSW#1{\put(-0.48,-0.52){\makebox(0,0)[tl]{$#1$}}\end{picture}}
\def\flS{\begin{picture}(0,0)
   \put(    0,-0.2 ){\vector( 0,-1){0.60}}
   \@ifstar{\@flS}{\@@flS}}
\def\@flS  #1{\put(-0.03,-0.5 ){\makebox(0,0)[ r]{$#1$}}\end{picture}}
\def\@@flS #1{\put( 0.03,-0.5 ){\makebox(0,0)[ l]{$#1$}}\end{picture}}
\def\flSE{\begin{picture}(0,0)
   \put( 0.18,-0.18){\vector( 1,-1){0.64}}
   \@ifstar{\@flSE}{\@@flSE}}
\def\@flSE #1{\put( 0.48,-0.52){\makebox(0,0)[tr]{$#1$}}\end{picture}}
\def\@@flSE#1{\put( 0.52,-0.48){\makebox(0,0)[bl]{$#1$}}\end{picture}}
\def\capsa(#1,#2)#3{\put(#1,#2){\makebox(0,0){$#3$}}}
\def\indiag{\@ifnextchar [{\@indiag}{\@indiag[15ex]}}
\def\@indiag[#1](#2,#3){\begingroup
   \setlength{\unitlength}{#1}
   \medskip
   \begin{center}
   \begin{picture}(#2,#3)}
\def\exdiag{\end{picture}
   \end{center}
   \medskip
   \endgroup}
\makeatother


\begin{titlepage}

\thispagestyle{empty}

\begin{flushright}
\hfill{CERN-PH-TH 2007/144} \\
\end{flushright}

\vspace{20pt}

\begin{center}

{ \Large{\bf On the Consistency of Coset Space Dimensional Reduction
}}

\vspace{35pt}

{\bf A.~Chatzistavrakidis}$^{1,2}$,  {\bf P.~Manousselis}$^{2,3}$,
{\bf N.~Prezas}$^4$ {\bf and} {\bf G.~Zoupanos}$^2$ \vspace{20pt}

$^1$ {\it Institute of Nuclear Physics,\\
NCSR  DEMOKRITOS,\\
GR-15310 Athens, Greece}\\
 \vspace{5pt}

$^2${\it Physics Department, National Technical University of Athens, \\
GR-15780 Zografou Campus, Athens, Greece} \\
\vspace{5pt}

$^3${\it Department of Engineering Sciences, University of Patras,\\
GR-26110 Patras, Greece}\\ \vspace{5pt}

$^4${\it CERN PH-TH,\\
1211 Geneva, Switzerland}\\
\vspace{5pt}

\

Email: {\tt cthan@mail.ntua.gr,  pman@central.ntua.gr, george.zoupanos@cern.ch,
nikolaos.prezas@cern.ch}

\vspace{35pt}

{ABSTRACT}
\end{center}
\vspace{5pt}

In this letter we consider higher-dimensional Yang--Mills theories and examine their
consistent coset space dimensional reduction. Utilizing a suitable ansatz
and imposing a simple set of
constraints we determine the four-dimensional gauge theory obtained
from the reduction of both the higher-dimensional Lagrangian and the
corresponding equations of motion. The two reductions yield equivalent results
and hence they constitute an example of a consistent truncation.


\vspace{20pt}


\vspace{20pt}

\end{titlepage}

\newpage

\baselineskip 6 mm

\section{Introduction}

The first attempt to unify interactions
through higher dimensions and dimensional reduction  
was  made by
Nordstr\"om \cite{Nordstrom:1914fn,Nordstrom:1988fi} even earlier than
the celebrated work of Kaluza and Klein \cite{Kaluza:1921tu,Klein:1926tv}.
The Kaluza--Klein proposal was to unify gravity
and electromagnetism by considering five-dimensional gravity 
dimensionally reduced on a circle
down to four dimensions. 
Compactification of higher-dimensional gravity was studied further by
Pauli \cite{pauli;53}, who assumed a  2-sphere as the extra-dimensional space,
and later by DeWitt \cite{dewitt;63}. One implication of considering
internal spaces with non-abelian isometry 
is that Yang--Mills fields appear naturally in lower dimensions\footnote{ A nice
review of these developments can be found in
ref.~\cite{Pons:2006vz}.}. As a result, a revival of 
interest in the Kaluza--Klein proposal was  inspired by the hope
 that a geometrical unification of gravity
with the other observed non-abelian gauge interactions could be achieved \cite{Kerner68}.

However, this ambitious program met also serious obstacles. The most
serious, plaguing all attempts of constructing realistic theories, was the inability to
obtain chiral fermions in four dimensions \cite{Witten:1981me}.
Fortunately, there is a very
interesting resolution to this problem achieved by adding Yang--Mills fields to
the original higher-dimensional theory
\cite{Manton:1981es, Chapline:1982wy}.
Introducing non-Abelian gauge fields in higher dimensions is also
welcome for another reason; it provides a potential
unification of the low-energy gauge interactions as well as of the gauge
and the Higgs fields. Concerning the latter we should recall that the
celebrated Standard Model  of elementary particle physics
has  obvious limitations due to the
presence of a plethora of free parameters mostly related to the
ad-hoc introduction of the Higgs and Yukawa sectors in the theory.

The Coset Space Dimensional Reduction (CSDR)
 \cite{Witten:1976ck,Forgacs:1979zs, Kapetanakis:1992hf} was suggesting from
the beginning that a unification of the gauge and Higgs sectors can
be achieved using higher dimensions. In the CSDR scheme one assumes that
the space-time is of the form $M^D = M^4 \times S/R$ with $S/R$ being a
homogeneous coset space. Then a gauge theory with gauge group $G$
defined on $M^D$ can be dimensionally reduced to $M^4$ in an elegant
way by employing the symmetries of $S/R$. The resulting
four-dimensional gauge group is a subgroup of $G$ and the
four-dimensional gauge and Higgs fields are simply the surviving
components of the gauge fields of the pure higher-dimensional gauge theory.

Upon introducing  fermions  \cite{Manton:1981es} the
four-dimensional Yukawa and fermion-gauge interactions  find 
a unified description in the gauge interactions of the
higher-dimensional theory. In this vein, another appealing feature of the CSDR
is the observation that utilizing non-symmetric coset spaces
leads to softly broken supersymmetric theories
\cite{Manousselis:2001re}. Let us finally mention that
a first step towards embedding the CSDR
in string theory was taken in
ref.~\cite{Manousselis:2005xa} where it was shown that six-dimensional
nearly-K\"ahler  coset spaces provide internal spaces for
supersymmetric heterotic string compactifications in the presence of
fluxes and condensates.

An important point, which has been raised by a number of authors
(e.g.~\cite{Cvetic:2003jy}), concerns the consistency of the various
reduction schemes. It is therefore natural to ask if the CSDR is consistent.
A consistent reduction scheme of higher-dimensional theories over a unimodular group
manifold $S$ was discovered by Scherk and Schwarz 
using the ansatz that the metric and the various spacetime fields
are invariant under either the left or the right translations of the group 
\cite{Scherk:1979zr}. A similar ansatz was employed in ref.~\cite{Surridge:1987ws}
for the reduction on a coset space
$S/R$ and one of the goals of this work was to obtain the CSDR constraints from the
Einstein--Yang--Mills equations of motion. It should be emphasized  that
there is no known recipe for obtaining consistent coset
reductions neither a systematic understanding of the conditions under
which such a reduction occurs. In the very few existing examples of
consistent coset reductions there is very little conceptual
understanding of the underlying reasons for the consistency
but  the detailed calculation
itself. Some examples can be found in refs.~\cite{Cvetic:2003jy,
Gibbons:2003gp} and in the references therein.

Recently the concepts of
truncation and  reduction as well as the criteria for consistency were re-examined and analyzed
 \cite{Pons:2003ka, Pons:2006vz}. The upshot is that
if we consider a theory described by a Lagrangian ${\mathcal L}$
as the starting point, then, for a certain number of dimensions of
space-time we can perform a truncation of  ${\mathcal L}$
by essentially two
methods:

{\bf (a)} first type: ordinary Kaluza--Klein dimensional reduction
where only the massless modes are kept,

{\bf (b)} second type: by introducing constraints that
reduce the number of independent fields -- or field components --
defining the theory.

These two procedures are usually applied simultaneously and constitute
the concept of dimensional reduction. In both cases we are performing
a {\sl truncation} in the field content of the theory, either
because Kaluza--Klein modes are eliminated in the dimensional
reduction or because some field components  are rendered redundant due to
the presence of constraints.
The  {\sl consistency} of such a truncation
 is decided by  
examining whether solutions of the equations of motion of  the
reduced Lagrangian ${\mathcal L}_R$ are still solutions of the
equations of motion of the original Lagrangian  ${\mathcal L}$. This
property is expressed graphically as the commutativity of the
following diagram: \indiag(1,1) \capsa(0,0){ \frac{\delta{\mathcal
L}}{\delta \Phi}\!=0 } \capsa(0,1){{\mathcal L}}
\capsa(0,1){\flS{{\rm e.o.m.}}} \capsa(0,0){\flE{{\rm Red}.}}
\capsa(0,1){\qquad\quad\flE{{\rm Red}.}} \capsa(1,0){\qquad\ \qquad
\qquad\big(\frac{\delta{\mathcal L}}{\delta \Phi}
\big)_{\!_{\!R}}\!=0 \Leftrightarrow \frac{\delta{\mathcal
L}_{\!R}}{\delta \Phi}=0 } \capsa(1.5,1){{\mathcal L}_R}
\capsa(1.5,1){\flS{{\rm e.o.m.}}} \exdiag

A proper definition is the following: {\sl a truncation is said to
be consistent when its implementation at the level of the
variational principle agrees with its implementation
at the level of the equations
of motion}, i.e., if these operations commute: first truncate the
Lagrangian and then obtain the equations of motion (e.o.m.) or
first obtain the e.o.m. and then truncate them.

Concerning truncations of the second type, i.e.~driven by the introduction of
constraints, the outcome of the analysis of \cite{Pons:2003ka, Pons:2006vz}
is that a mechanism parallel to that of the
Dirac--Bergmann theory of constrained systems applies. In general, the
presence of secondary constraints -- dynamically derived from the
original (primary) ones -- is a typical obstruction to the consistency of
the truncation. In this letter we prove that the coset space dimensional reduction
of \cite{Forgacs:1979zs} is an example of a truncation of the second type
and that it  is consistent in the sense explained earlier. In
particular no new constrains are needed for the reduction of the
higher-dimensional equations of motion besides the ones necessary for
the proper reduction of the Lagrangian.

\section{Reduction on coset spaces: brief reminder}

\subsection{Geometry of coset spaces}

The geometry of coset spaces $S/R$ relevant for our purposes is presented
in refs.~\cite{Castellani:1983tb, Castellani:1999fz}. Let the
coordinates of the Lie group $S$ be $(y^{a}, z^{i})$ with
$y^{a}$ being the coset coordinates and $z^{i}$ being the coordinates of the $R$
subgroup. Then a group element $s \in S$ can be represented as $s
\sim e^{y^{a}Q_{a}} e^{z^{i}Q_{i}}$ and a coset representative is
$L(y) = e^{y^{a}Q_{a}}$. The
Maurer--Cartan 1-form is defined by $e(y)=L^{-1}(y)dL$ and is the
analogue of the left-invariant 1-form on a Lie group $S$. It takes values in
$Lie(S)$, i.e.~the Lie algebra of $S$:
\begin{equation}
e(y) = e^{A}Q_{A}= e^{a}Q_{a} + e^{i}Q_{i},
\end{equation}
where $A$ is a group index, $e^{a}$ is the coframe and $e^{i}$ is the
$R$-connection. The latter can be expanded 
in coset vielbeins as $e^i=e^i_a(y) e^a$.
The exterior
derivative of the Maurer--Cartan 1-form is
\begin{equation}
de = d(L^{-1}dL) = - e \wedge e = - [e,e],
\end{equation}
from which we can easily prove that
\begin{equation}
de^{A} = - \frac{1}{2}f^{A}_{BC} e^{B} \wedge e^{C}.\label{mcg}
\end{equation}

We will assume, for reasons analyzed in detail in
ref.~\cite{Castellani:1983tb}, that the coset is reductive. That
means that the commutation relations obeyed by the generators of $S$
are not the most general ones but they take the form
\begin{eqnarray}
\left[ Q_{i}, Q_{j} \right] &=& f_{ij}^{k} Q_{k}, \nonumber \\
\left[Q_{i}, Q_{a}\right] &=& f_{ia}^{b}Q_{b}, \nonumber \\
\left[Q_{a},Q_{b}\right]&=& f_{ab}^{c} Q_{c} + f_{ab}^{i}
Q_{i},
\end{eqnarray}
implying that  $f_{bi}^{j}=0$.
Now (\ref{mcg}) can be written as  
\begin{eqnarray}
de^{a} = - \frac{1}{2}f^{a}_{ bc}e^{b} \wedge e^{c} -
f^{a}_{ bi}e^{b} \wedge e^{i}, \label{first} \\
de^{i} = -\frac{1}{2} f^{i}_{ab}e^{a} \wedge e^{b} -
\frac{1}{2}f^{i}_{ jk}e^{j} \wedge e^{k}
\end{eqnarray}
and from eq.~(\ref{first}) we can obtain the Maurer--Cartan equations for the
coset vielbeins
\begin{equation}\label{second}
de^{a} = - \frac{1}{2}C^{a}_{ bc}(y) e^{b} \wedge e^{c}, \;\;\;
C^{a}_{ bc} = f^{a}_{ bc} - 2e^{i}_{[b}f^{a}_{c]i}.
\end{equation}

It is straightforward to verify the following identities that we will
frequently use in the ensuing
\begin{eqnarray}
&&e^{a} \wedge \ast_{d} e^{b} = \delta^{ab} vol_{d},\\
&&d \ast_{d} e^{a} =  C^{b}_{ba}(y)
vol_{d},\label{2ui}\\
&&e^{a} \wedge \ast_{d} (e^{b} \wedge e^{c}) = \delta^{ac} \ast_{d}
e^{b} - \delta^{ab} \ast_{d} e^{c}, \label{3ui}\\
&&d\ast_{d}(e^{a} \wedge e^{b}) = (\ast e^a) C^c_{cb} -(\ast e^b)
C^c_{ca}-C^c_{ab} (\ast e^c)\label{4ui}.
\end{eqnarray}
In these formulas $vol_d$ is the volume form of the coset. Notice that
the tangent coset indices $a,b,\ldots$ are raised and lowered
by $\delta_{ab}$ and hence upper and lower such indices are equivalent.

Finally, let us mention that there exist $S$-invariant metrics on $S/R$. An
example is
\begin{equation}
g(y) = \gamma_{a b}e^{a}(y)e^{b}(y),
\end{equation}
where $\gamma_{a b}$ is the Killing metric of the group restricted
on the coset $S/R$. The Killing vectors associated with the left
isometry group $S$ will be denoted by $X_{I}$.

\subsection{Reduction on group manifolds and coset spaces}

Before we describe the Coset Space Dimensional Reduction (CSDR) scheme, let us
recall that the ansatz for the Scherk--Schwartz  reduction \cite{Scherk:1979zr} of 
higher-dimensional gauge field $A$ on a group manifold $S$ is
\begin{equation}\label{SS}
A=A_{\mu}dx^{\mu} + A_{I}(x)e^{I}(y),
\end{equation}
with $I=1,\ldots,{\rm dim} S$ and $e^{I}$ being the left-invariant
1-forms. Hence, in this type of reduction on group manifolds we keep only the
$G_{L}$ singlets under the full isometry group  $G_{L}
\times G_{R}$ of $G$. This truncation can be described by the following
invariance condition:
\begin{equation}
{\cal L}_{X^{I}}A = 0,
\end{equation}
with $X^{I}$ being the Killing vectors  dual to the right-invariant
1-forms\footnote{Recall that the right-invariant vector fields generate
left-translations.}.
The Scherk--Schwarz reduction of the metric is performed by enforcing a similar
invariance condition
\begin{equation}\label{InvarianceCondition}
{\cal L}_{X^{I}}g_{MN} =0.
\end{equation}

In the CSDR one allows a generalized invariance condition
\begin{equation}\label{GSC}
{\cal L}_{X^{I}}A = DW_{I},
\end{equation}
with $W_{I}$ being a gauge transformation parameter associated to
the Killing vector $X_{I}$. The invariance condition for the
reduction of the metric is the same as in the group manifold
case since the metric is gauge singlet. The generalized
invariance condition
\begin{equation}\label{21}
{\cal L}_{X^{I}}A = i_{X^{I}}dA + di_{X^{I}}A = DW_{I} = dW_{I} +
[A,W],
\end{equation}
together with the consistency condition
\begin{equation}\label{22}
\left[{\cal L}_{X^{I}}, {\cal L}_{X^{J}} \right] = {\cal
L}_{[X^{I},X^{J}]},
\end{equation}
impose constraints on the gauge field. The detailed analysis of the
constraints (\ref{21}) and (\ref{22}) given in
refs.\cite{Forgacs:1979zs,Kapetanakis:1992hf} provides us with the
four-dimensional unconstrained fields as well as with the residual gauge
invariance of the lower-dimensional theory.

\section{CSDR of  Yang--Mills theory}

\subsection{Gravity background}

The question of compatibility of the CSDR constraints with the
equations of motion was originally raised in \cite{Surridge:1987ws}. In this
letter we revisit  this question in light of the recent analysis
of \cite{Pons:2003ka, Pons:2006vz}.

We begin by examining a
$D$-dimensional Einstein--Yang--Mills Lagrangian
\begin{equation}
{\cal L} = \hat{R} \ast_{D} {\bf 1} - \frac{1}{2}{\rm Tr}\hat{F}_{(2)}
\wedge \ast_{D} \hat{F}_{(2)} - \lambda_{(D)} \ast_{D} {\bf 1},
\label{hdaction}
\end{equation}
where
$$\hat{F}_{(2)}=d\hat{A}_{(1)} + \hat{A}_{(1)} \wedge \hat{A}_{(1)},$$
is a non-Abelian gauge field strength taking values in the Lie algebra of a group $G$.
$\hat{R}$ is the curvature scalar and $\lambda_{(D)}$ is the
cosmological constant in $D$-dimensions. We distinguish the
higher-dimensional fields from the four-dimensional ones by putting hats
on them. The equations of motion are
\begin{eqnarray}
  &&\hat{R}_{MN}  = \frac{1}{2}Tr(\hat{F}^{2}_{MN} - \frac{1}{8}\hat{F}^{2}\hat{g}_{MN}) +
  \frac{1}{4}\lambda_{(D)}\hat{g}_{MN}, \label{system} \\
  &&\hat{D}(\ast_{D} \hat{F}_{(2)}) = 0.\label{system1}
\end{eqnarray}

We will perform a reduction on a spacetime
 of the form $M_{4} \times S/R$ where $M_{4}$ is the
four-dimensional Minkowski spacetime and $S/R$ is a compact coset
space of dimension $d=D-4$.
This background is described by a metric of the form\footnote{
We denote tangent spacetime indices by $m,n,\ldots$ and
curved ones by $\mu,\nu,\ldots$, tangent coset 
indices by $a,b,\ldots$  and curved ones by  $\alpha,\beta,\ldots$, 
while for the full group $S$  
tangent indices are denoted by $I,J,\ldots$ and curved ones by  $A,B,\ldots$. Indices
on the subgroup R are denoted by $i,j,\ldots$.Finally, the spacetime coordinates
are $x^\mu$ while those of the coset space are $y^\alpha$.}
\begin{equation}\label{InvariantMetric}
g_{(D)} = \eta_{mn}e^{m}e^{n} + \gamma_{ab}e^{a}e^{b},
\end{equation}
where $\eta_{mn}$ is the flat Minkowski metric and $\gamma_{ab}$ is
an $S$-invariant metric on the coset $S/R$ satisfying
(\ref{InvarianceCondition}). From the metric
(\ref{InvariantMetric}) and using the vielbeins  constructed from the
Maurer--Cartan 1-form on the coset $S/R$ we can compute
the Ricci curvature and Ricci scalar. The Einstein equations (\ref{system}) 
take the form
\begin{equation}
\frac{1}{4} Tr F_{ab}F^{ab} = \lambda_{(D)},
\end{equation}
and
\begin{equation}
TrF^{2}_{ab} = R_{(d)}\gamma_{ab},
\end{equation}
where $\lambda_{D}$ is the higher-dimensional cosmological constant
and $R_{(d)}$ is the scalar curvature of the internal space.
Let us note  that the complete elimination  of the Kaluza--Klein gauge
fields as well as  of the scalar moduli is a well-known 
consistent truncation of gravity on coset spaces
\cite{Pons:2003ka}. We plan to examine a more general reduction scheme of
gravity on cosets in future work \cite{cmpz}.

\subsection{Reduction of the action}
The reduction ansatz for the gauge fields is
\begin{equation}
\hat{A}^{\tilde{I}}(x,y) = A^{\tilde{I}}(x) +
\chi^{\tilde{I}}_{\alpha}(x,y)dy^{\alpha},
\end{equation}
where
\begin{equation}
\chi^{\tilde{I}}_{\alpha}(x,y) =
\phi^{\tilde{I}}_{A}(x)e^{A}_{\alpha}(y)
\end{equation}
and $\tilde{I}$ is a Lie algebra index $\tilde{I}= 1, \ldots, {\rm dim}
G$. 
This formally looks exactly the same as the reduction ansatz on
group manifolds. The objects $\phi_{A}^{\tilde I}(x)$ take values in the
Lie algebra of $G$ 
 and are coordinate scalars in four dimensions that
can be interpreted as Higgs fields. The ansatz solves the
generalized symmetry condition (\ref{GSC}) provided we can embed $R$
in $G$ as explained in
refs.~\cite{Forgacs:1979zs,Kapetanakis:1992hf}.

Next we substitute the above ansatz into the action
and subsequently into the equations of motion. The 
gauge field can be written as
\begin{equation}\label{YMCSDR}
\hat{A}^{\tilde{I}}=A^{\tilde{I}} + \phi^{\tilde{I}}_{A}e^{A},
\end{equation}
and the corresponding field strength is
\begin{equation}
\hat{F}^{\tilde{I}} = d\hat{A}^{\tilde{I}} +
\frac{1}{2}f^{\tilde{I}}_{\tilde{J} \tilde{K}}
\hat{A}^{\tilde{J}} \wedge A^{\tilde{K}}
\end{equation}
with $f^{\tilde{I}}_{\tilde{J} \tilde{K}}$ being the structure constants
of the higher-dimensional gauge group $G$.

Using (\ref{mcg}) 
we compute
\begin{equation}\label{Sans}
\hat{F}^{\tilde{I}} = F^{\tilde{I}} + D\phi^{\tilde{I}}_{A} \wedge
e^{A} - \frac{1}{2} F^{\tilde{I}}_{AB} e^{A} \wedge e^{B},
\end{equation}
where
\begin{eqnarray}
F^{\tilde{I}} &=& dA^{ \tilde{I} } + \frac{1}{2} f^{ \tilde{I} }_{
\tilde{J} \tilde{K}} A^{\tilde{J}} \wedge A^{\tilde{K}},\\
D\phi^{\tilde{I}}_{A} &=& d\phi^{\tilde{I}}_{A} + f^{\tilde{I}}_{
\tilde{J} \tilde{K}} A^{\tilde{J}} \phi^{\tilde{K}}_{A}, \\
F^{\tilde{I}}_{AB} &=& f^{C}_{A B} \phi^{\tilde{I}}_{C} -
[\phi_{A} , \phi_{B}]^{\tilde{I}}.
\end{eqnarray}
Note that we obtain (\ref{Sans}), which admits gauge invariant constraints, 
provided we use the ansatz (\ref{YMCSDR}).
Had we used an ansatz of the form $\hat{A}^{\tilde{I}}(x,y) =
A^{\tilde{I}}(x) + \phi^{\tilde{I}}_{a}(x)e^{a}$, then instead 
of structure constants $f^{A}_{BC}$ we would have obtained the  tensors 
$C^{a}_{bc}(y)$ which are functions of $y$.

Next we will  insert eq.~(\ref{Sans}) in (\ref{hdaction})
therefore obtaining the effective four-dimensional action and the CSDR
constraints. 
To compute the higher-dimensional Yang--Mills action
\begin{equation}\label{HDAction}
S = -\frac{1}{2} \int Tr \hat{F} \wedge \ast_{D} \hat{F},
\end{equation}
we need to Hodge-dualize  eq.~(\ref{Sans}) to
\begin{equation}\label{SansDual}
\ast_{D}\hat{F}^{\tilde{I}} = \ast_{4}F^{\tilde{I}} \wedge vol_{d} +
\ast_{4}D\phi_{A}^{\tilde{I}} \wedge \ast_{d}e^{A} -
\frac{1}{2}F_{AB} vol_{4} \wedge \ast_{d}(e^{A} \wedge e^{B}).
\end{equation}
In this formula $vol_4$ is the volume form of the four-dimensional spacetime
while $\ast_4$ and $\ast_d$ denote the Hodge duality operation in spacetime
and the coset respectively.

Inserting (\ref{Sans}) and (\ref{SansDual}) in (\ref{HDAction})
we obtain the four-dimensional Lagrangian
\begin{equation}
{\cal L} =  -\frac{1}{2} Tr F \wedge \ast_{4} F + \frac{1}{2} Tr
D\phi_{a} \wedge \ast_{4}D \phi _{a} - \frac{1}{4} F_{ab}F^{ab}
vol_{4},\label{4dl}
\end{equation}
provided we impose the CSDR constraints:
\begin{equation}\label{constraints}
D\phi^{\tilde{I}}_{i} = F^{\tilde{I}}_{ai} = F^{\tilde{I}}_{ij} =0.
\end{equation}
Explicitly
the constraints (\ref{constraints}) read\footnote{In order to avoid
cluttering we will occasionally suppress the Lie algebra indices.}
\begin{eqnarray}
  F_{ib} &=& f_{ib}^{ \ \ c}\phi_{c} -[\phi_{i}, \phi_{b}]=0, \label{417} \\
  F_{ij}&=& f_{ij}^{ \ \ k}\phi_{k} - [\phi_{i},\phi_{j}]=0, \label{418}
\end{eqnarray}
as well as
\begin{equation}\label{419}
[A_{\mu}, \phi_{i}] = 0.
\end{equation}
The last equation comes from $D\phi^{\tilde{I}}_{i}$ assuming
constant values for the scalars $\phi^{\tilde{I}}_{i}$ as dictated
by Lorentz invariance. Notice that $F_{ab}$ depends on $\phi_i$
and in (\ref{4dl}) the latter are set to their constant values which are 
determined by (\ref{418}).

\subsection{Four-dimensional gauge group and spectrum}

At this point let us briefly explain how the above constraints
determine the four-dimensional gauge group and the spectrum.
Eq.~(\ref{418}) implies that the extra components
$\phi_{i}$  are generators of $R$. Furthermore we assume 
that $R$ is a Lie subgroup of $G$.
Then from (\ref{419}) we conclude that the four-dimensional gauge fields
$A_\mu$ have to commute
with the generators  of the $R$ subgroup of $G$. In other words, the
gauge group in four dimensions $H$ is the centralizer of the embedding
of $R$ in $G$: $H=C_{G}(R)$. 

The fields $\phi_{a}(x)$  are
scalars in four dimensions. From (\ref{417}) we see that 
these fields transform under $R$ as a
vector $v$
\begin{equation}
S \supset R, \;\;\;
adjS = adjR+v,
\end{equation}
and furthermore that  the $\phi_{a}(x)$ are
intertwining operators connecting induced representations of $R$
acting on $G$ and $S/R$. This implies, exploiting Schur's lemma,
that the transformation properties of the fields $\phi_{a}$ under
$H$ can be deduced if we express the adjoint representation of $G$ in
terms of $R \times H$ :
\begin{equation}
G \supset R,\;\;\;
 adjG =(adjR,1)+(1,adjH)+\sum(r_{i},h_{i}).
\end{equation}
Then if $v=\sum s_{i}$, where each $s_{i}$ is an irreducible
representation of $R$, we obtain an $h_{i}$ multiplet of scalars $\phi_a$ for every
pair $(r_{i},s_{i})$ with $r_{i}$ and $s_{i}$ being identical
irreducible representations of $R$.

\subsection{Reduction of the equations of motion}

Next we examine the reduction of the higher-dimensional equations of
motion. Since we have imposed constraints 
the consistency of this reduction 
 is not guaranteed by the consistency
of the reduction of the higher-dimensional Lagrangian \cite{Pons:2003ka}.  
Our final result will be that the diagram presented in
the introduction is commutative, thereby proving the consistency of the CSDR
scheme. 

The higher-dimensional Yang--Mills equation is
\begin{equation}
\hat{D} \ast_{D} \hat{F}^{\tilde{I}} = \label{YMEOMS}
\hat{d} \ast_{D} \hat{F}^{\tilde{I}} + f^{\tilde{I}}_{\tilde{J}
\tilde{K}} \hat{A}^{\tilde{J}} \wedge \ast_{D}\hat{F}^{\tilde{K}} =
0.
\end{equation}
Substituting the reduction ansatz (\ref{YMCSDR}, \ref{SansDual})
into (\ref{YMEOMS}) we obtain
\begin{equation}\label{4dimym}
D \ast_{4} F^{\tilde{I}} \wedge vol_{d} -f^{\tilde{I}}_{
\tilde{J} \tilde{K}}\phi^{\tilde{J}}_{A}
\ast_{4}D\phi^{\tilde{K}}_{B} \wedge e^{A} \wedge \ast_{d} e^{B}
- \ast_{4}D\phi^{\tilde{I}}_{A} \wedge d\ast_{d}e^{A}
=0
\end{equation}
and
\begin{equation}\label{4dimsc}
D \ast_{4}D\phi^{\tilde{I}}_{A} \wedge \ast_{d} e^{A} - \frac{1}{2}
f^{\tilde{I}}_{ \ \ \tilde{J}
\tilde{K}}\phi^{\tilde{J}}_{A}F^{\tilde{K}}_{BC}vol_{4} \wedge e^{A}
\wedge \ast_{d}(e^{B} \wedge e^{C})
- \frac{1}{2}F^{\tilde{I}}_{BC}vol_{4} \wedge d \ast_{d}(e^{B}
\wedge e^{C})=0.
\end{equation}

Using these relations and imposing the same constraints
(\ref{constraints}) 
we used earlier in the reduction of the Lagrangian,
we obtain the equations of motion for the four-dimensional gauge fields
and scalars:
\begin{eqnarray}
D \ast_{4} F^{\tilde{I}} &=&  f^{\tilde{I}}_{  \tilde{J}
\tilde{K}}\phi^{\tilde{J}}_{a} \ast_{4} D\phi^{\tilde{K}}_{a}, \label{4YM}\\
D \ast_{4} D \phi^{\tilde{I}}_{a} &=& -(f^{\tilde{I}}_{\tilde{J}
\tilde{K}}\phi^{\tilde{J}}_{c}F^{\tilde{K}}_{ca} +\frac{1}{2}
F_{bc}f_{bca}) vol_4.\label{4Sclr}
\end{eqnarray}
These are the equations of motion  for a Yang--Mills theory coupled to
charged scalars with a non-trivial potential as described by
the Lagrangian (\ref{4dl}). Therefore the consistency of the truncation
is verified.

In order to deduce (\ref{4YM}) from (\ref{4dimym}) we have to show that
the extra terms
\begin{equation}
-f^{\tilde{I}}_{ \ \ \tilde{J} \tilde{K}}\phi^{\tilde{J}}_{i}
\ast_{4}D\phi^{\tilde{K}}_{b} \wedge e^{i}_{a}e^{a} \wedge \ast_{d}
e^{b} - \ast_{4}D\phi^{\tilde{I}}_{a} \wedge d\ast_{d}e^{a}\label{et1}
\end{equation}
cancel out. 
First we notice that if $S$ is unimodular, i.e.~it has traceless structure constants $f^{A}_{AB}=0$, and the coset is reductive we have $f^a_{ab}=0$. Then from
 (\ref{second}) and (\ref{2ui}) we find  
$ d\ast_d e^a=-e^i_b f^b_{ai} vol_d$.
 Subsequently  (\ref{et1}) becomes
 \begin{equation}
\ast_4 \Big(f^b_{ai} D\phi_a^{\tilde I}- 
 f^{\tilde{I}}_{ \ \ \tilde{J} \tilde{K}} \phi^{\tilde{J}}_{i}
D\phi^{\tilde{K}}_{b}\Big) e^i_b \wedge vol_d = \ast_4 (DF_{ib}^{\tilde I}) e^{i}_b \wedge vol_d=0
\end{equation}
 where we have used the constraints $D\phi_i^{\tilde I}=0$ and
 $F_{ib}=0$.
  
In order to derive  (\ref{4Sclr}) from (\ref{4dimsc}) we write the terms
 \begin{equation}\label{pd}
- \frac{1}{2} f^{\tilde{I}}_{\tilde{J}
\tilde{K}}\phi^{\tilde{J}}_{i}F^{\tilde{K}}_{bc}vol_{4} \wedge e^{i}
\wedge \ast_{d}(e^{b} \wedge e^{c}) -
\frac{1}{2}F^{\tilde{I}}_{bc}vol_{4} \wedge d \ast_{d}(e^{b} \wedge
e^{c})
\end{equation}
and show that they equal the second term at the right-hand side of (\ref{4Sclr}).
Using (\ref{second}), (\ref{3ui}) and (\ref{4ui})
we can re-write them as
\begin{eqnarray}
vol_4 &\wedge& \ast e^c \Big[
[\phi_{i}, F_{bc}]^{\tilde I}e^{ib}
- F_{ab}^{\tilde I} e^i_a f^c_{bi}  + F_{cb}^{\tilde I} e^i_a f^a_{bi}  +\frac{1}{2}
F_{ab}^{\tilde I} f^c_{ab} \Big]=\nonumber\\
 vol_4 &\wedge&  \ast e^c  \Big[e^i_b \big(
[\phi_{i}, F_{bc}]^{\tilde I}
- (F_{ba}^{\tilde I} f^c_{ai}  + F_{ca}^{\tilde I} f^b_{ai})\big)+\frac{1}{2}
F_{ab}^{\tilde I} f^c_{ab}\Big] . \nonumber
\end{eqnarray}
The last term in this formula contributes to the equation of motion
(\ref{4Sclr}) while the rest cancel out. 
The cancellation is due to the identity
\begin{equation}\label{F}
[\phi_{i}, F_{ab}]^{\tilde I} = f_{ia}^{c}F_{cb}^{\tilde I} + f_{ib}^{c}F_{ac}^{\tilde I}
\end{equation}
that holds as a consequence  of the constraint $[\phi_{i},
\phi_{a}]^{\tilde I} = f_{ib}^{ \ c}\phi_{c}^{\tilde I}$ and the definition of $F_{ab}^{\tilde I} = f_{ab}^{c}\phi_{c}^{\tilde I} +f_{ab}^{i}\phi_{i}^{\tilde I} - [\phi_{a}, \phi_{b}]^{\tilde I}$,
provided that we use the Jacobi identity and assuming that 
the structure constants $f_{ABC}$ of the Lie group $S$ are completely
antisymmetric and that the coset $S/R$ is reductive.

\subsection{Bianchi identities}

To conclude our analysis we examine the higher-dimensional Bianchi
Identity
\begin{equation}
\hat{D} \hat{F}^{\tilde{I}} = 
\hat{d}\hat{F}^{\tilde{I}} +f^{\tilde{I}}_{ \tilde{J}
\tilde{K}}\hat{A}^{\tilde{J}} \wedge \hat{F}^{\tilde{K}} = 0.\label{Bianchi}
\end{equation}
Substituting the  reduction ansatz (\ref{YMCSDR}) and (\ref{Sans}) into
eq.~(\ref{Bianchi}) we obtain
\begin{eqnarray}
&&dF^{\tilde{I}} +f^{\tilde{I}}_{\tilde{J} \tilde{K}}
A^{\tilde{J}} \wedge F^{\tilde{K}} = 0, \\
&&d D\phi^{\tilde{I}}_{A} \wedge e^{A} + f^{\tilde{I}}_{ 
\tilde{J} \tilde{K}} A^{\tilde{J}} \wedge D\phi^{\tilde{K}}_{A}
\wedge e^{A} - f^{\tilde{I}}_{ \tilde{J}
\tilde{K}}\phi^{\tilde{J}}_{A}F^{\tilde{K}} \wedge e^{A} = 0,\\
&&DF^{\tilde{I}}_{BC} = f^{A}_{  BC}D\phi^{\tilde{I}}_{A} -
2f^{\tilde{I}}_{ \tilde{J}
\tilde{K}}\phi^{\tilde{J}}_{B}D\phi^{\tilde{K}}_{C}, \\
&&\frac{1}{2}(F^{\tilde{I}}_{AB}f^{A}_{  DC}-f^{\tilde{I}}_{
\tilde{J} \tilde{K}}\phi^{\tilde{J}}_{B}F^{\tilde{K}}_{DC})e^{D}
\wedge e^{C} \wedge e^{B} = 0.
\end{eqnarray}
These equations, after
imposing the  constraints (\ref{constraints}) and utilizing the Jacobi
identities for the structure constants, yield 
\begin{itemize}
\item the Bianchi identity for the four-dimensional
gauge field $DF =0$,
\item the Bianchi identity for the four-dimensional
scalars $D^{2}\phi = F \wedge \phi$.
\end{itemize}

\section{Conclusions}

We have studied a particular dimensional reduction scheme of
Yang--Mills theories with the
understanding that there are no general recipes for
consistent coset space reductions. Utilizing an ansatz inspired by
an example of a consistent group manifold reduction, 
we have examined to what extent a 
coset reduction can be consistent. We demonstrated that upon imposing
a simple set of constraints, which  we identified as the constraints
appearing naturally  in the CSDR reduction scheme, one can reduce
in a self-consistent manner both the higher-dimensional
Lagrangian and the corresponding equations of motion. The next and substantially
more involved
step is to include gravity fluctuations. Work in this direction
is currently in progress \cite{cmpz}.

\section*{Acknowledgements}
This work is supported by
the EPEAEK program "Pythagoras" and co-funded by the European Union (75\%)
and the Hellenic State (25\%). This work is also partially supported by
the NTUA program for basic research "Karatheodoris". PM is supported
by the Hellenic State Scholarship Foundation (I.K.Y.) and by the program
Pythagoras I (89194).

\end{document}